\documentclass[]{aastex631}

\usepackage{natbib}
\usepackage[flushleft]{threeparttable}

\shorttitle{Exoplanet classification}
\shortauthors{Prithiv and Alka.}

\graphicspath{{./}{figures/}}
\usepackage{csquotes}

\begin{document}

\title{Identification and Classification of Exoplanets using Machine Learning Techniques}

% \footnote{\href{blue}{prithivrajastro@gmail.com}}}

\author{Prithivraj G}
\email{prithivrajastro@gmail.com}
\author{Alka Kumari}
\email{Alka.Kumari@liverpool.ac.uk}

\begin{abstract}
NASA's Kepler Space Telescope has been instrumental in the task of finding the presence of exoplanets in our galaxy. This search has been supported by computational data analysis to identify exoplanets from the signals received by the Kepler telescope. In this paper, we consider building upon some existing work on exoplanet identification using residual networks for the data of the Kepler space telescope and its extended mission K2. 
%Our aim for this study is to identify and classify the exoplanet's light curves detected by the Kepler mission using various Machine Learning Algorithms. As the data is in bulk and noisy time-series data, manual computations are not viable to detect exoplanets. As a result, we can utilise machine learning methods to process this large amount of data and classify exoplanets. 
This paper aims to explore how deep learning algorithms can help in classifying the presence of exoplanets with less amount of data in one case and a more extensive variety of data in another. In addition to the standard CNN-based method, we propose a Siamese architecture that is particularly useful in addressing classification in a low-data scenario. The CNN and ResNet algorithms achieved an average accuracy of 68\% for three classes and 86\% for two-class classification. However, for both the three and two classes, the Siamese algorithm achieved 99\% accuracy.

\end{abstract}


\section{Introduction}
\paragraph*{} \quad Finding planets outside our solar system is challenging, and the first official exoplanet was detected in 1992, which was orbiting around the pulsar PSR B1257+12 (\cite{wolszczan1994confirmation}). At first, scientists used telescopes to examine exoplanets directly, but the most successful technique is the indirect detection methods such as the transit and radial velocity method. The first official space telescope dedicated only to the detection of exoplanets is Kepler Space Telescope (\cite{koch2010kepler}). The telescope was launched on March 7, 2009, and worked till November 15, 2018, for approximately nine years, resulting in the identification of more than 2,662 exoplanets outside our solar system. Such identification is based on analyzing large data to differentiate signals of candidate exoplanets from other signals.

%many of which might be potential planets for future life detection. The Kepler data covers until November 14, 2012, but we still have such a vast amount of data that will help us detect more new exoplanets in the future. 

\paragraph*{} \quad After the failure of the second and fourth wheels of the Kepler telescope four years into the mission, observations were temporarily suspended. As a result, scientists and engineers have decided to prolong the project with the K2 spacecraft. K2 is similar to Kepler but with a different field of view \cite{haas2014kepler}. This new field of view aids in viewing the other parts of the sky where we observe 100-degree square fields close to the ecliptics. It uses the Kepler spacecraft to look at the different parts of the sky for 80 days, with 4-5 days observing campaigns in a year. K2 will observe in both the northern and the southern hemispheres, and later, it covered ten times more sky area than the Kepler space mission. The K2 mission produced similar types of pixel-level and light curves data as produced by the Kepler mission using the same Photometer instrument, which is used to measure the light coming from a parent star. Consequently, a broad portion of the telescope is focused on detecting and investigating interesting objects such as supernovae, galaxies, stars, Pleiades, Neptune, comets and other heavenly bodies of the asteroid belt. It has found over 300 confirmed exoplanets in addition to 500 candidates (\cite{yu2018planetary})(\cite{mayo2018275}). Some of them can be habitable zone, while others are still waiting to be observed by future missions for further information.

\paragraph*{} \quad The Kepler mission was a big success in collecting bulk data by observing around 530,506 stars which yielded about 21 TB of publicly available data for such a data analysis process. This mission has helped us with the discovery of 2,662 confirmed exoplanets with 3,697 planet candidates from a set of 18,406 transit-like features detected on over 200,000 distinct stars. Apart from exoplanets, the mission also documented 61 supernova detection and enabled us to study and solve a few mysteries about it (\cite{cowen2014kepler}). For the Kepler data, the data is a total of 17 quarters (\cite{li2019kepler}), but for K2 the data is not over quarters but over a single field of view. One quarter is equal to three months or 90 days in a year. The observational period for the Kepler data is classified as long cadence (29.4 minutes) and short cadence (58.89 sec) (\cite{yang2018long}). 

% The first mission is to explore the structure of the Earth-like planets or Extra-solar planets (termed Exoplanets) near Habitable Zone (HZ) throughout our Milky Way galaxy. These exoplanets help us to view and observe various stages of planets and stars, and furthermore, the mission data helps us develop computational simulations of these stars and planets' formation in the early stages. 

% With this mission's contribution, we have managed to publish nearly 2,946 papers worldwide. The mission's data is used to explore different things to explain the most important questions about the working of our universe. (\cite{barentsen2018kepler}). 

\section{Machine Learning in Exoplanet Detection and the Role of Astronet-K2}
% \textcolor{red}{Role of ML in exoplanet detection (the headings in red colour are for our reference they will be removed after the paper correction)}
% \paragraph*{} \quad Machine learning is a branch of artificial intelligence that involves the development of algorithms and statistical models that enable computers to learn from and make predictions or decisions based on data without being explicitly programmed. 

\paragraph*{} \quad In recent years, machine learning has been increasingly applied to the field of exoplanet research to aid in the classification of exoplanets based on some observed attributes. Exoplanet classification refers to the process of categorizing exoplanets based on characteristics such as size, mass, composition, and orbital properties. Machine learning can be used to automate this process by training algorithms on large sets of data, such as those generated by telescopes and space missions, and then using those algorithms to classify new exoplanets based on their observed features. The common approach in machine learning-based classification of exoplanets involves using supervised learning techniques, where the algorithm is trained on a labelled dataset, i.e., a dataset where each exoplanet is assigned a specific class or category. The algorithm then uses this training data to learn patterns and relationships between the various properties of exoplanets and their assigned categories, which can then be used to predict the category of new exoplanets based on their observed properties.  

% Objects in telescopic images can be classified using machine learning. For example, the resolution of a telescopic image of a galaxy or a cluster of stars may not be sufficient to distinguish individual stars. Computer processing, on the other hand, can differentiate light from various stars in the population, much like low-resolution pictures may be enhanced to expose features not seen in the original. 

\paragraph*{} \quad In this context machine learning approaches use attributes based on observed patterns that change over time. Exoplanet hunting is one of them. Exoplanets are planets that orbit other stars. These worlds block a modest amount of light when they travel between their host stars and Earth. The amount of light that variates over time may be used to calculate the planet's size and orbit and this categorizes the detection of such bodies as exoplanets. Machine learning has been used to find several exoplanets (\cite{Malik_2021}), including a couple in multiple-planet systems where the signals are difficult to discriminate. Some stars are quite "active," emitting flares at irregular periods. Others have a changeable brightness, changing as they expand and compress. Computers are well-suited to detect these differences, which can be minute when compared to the massive quantity of data required to uncover them. Both existing and future observatories process terabytes of information on a daily basis. With so much data, determining what is relevant and what is not may be difficult, with the value varying depending on the scientific questions being answered. Astronomers can train computers to sort through the avalanche of data and pick out the crucial bits. Many existing and prospective observatories, like NASA's Transiting Exoplanet Survey Satellite (TESS), will collect even more data that will be beneficial in a variety of fields along with exoplanet detection. 

% Other machine learning techniques such as unsupervised learning and deep learning can also be applied to exoplanet classification, allowing for more complex patterns and relationships to be learned and utilized in the classification process. As an end result, machine learning will become increasingly crucial in the next years, and scientists are at the forefront of this progress.

% \textcolor{red}{Astronet-K2}
\paragraph*{} \quad For the classification of exoplanets, \cite{Shallue_2018} used the human-classified Kepler Threshold Crossing Events (TCEs) using Deep Learning algorithms involving a simple neural network and convolutional neural networks. Similarly, his work was extended by \cite{dattilo2019identifying} to uncover two new super-earths in K2 data. As the data is highly imbalanced in both the mission, the data has been preprocessed through a series of data analyses and preprocessing methods referred to as Astronet-K2 (\cite{dattilo2019identifying}), which we will be discussing later in this section. AstroNet-K2 is a neural network for identifying exoplanets in light curves and is indeed very good at classifying exoplanets and false candidates, with high accuracy. 
% The new exoplanet signatures were promptly identified by AstroNet-K2 and may investigate other parts of the galaxy in the same way, looking at stars that formed in various environments. 

% \textcolor{red}{what we are doing differently and contribution}
\begin{itemize}
 
     \item For this work, compared to the original data, we utilised fewer data from the Kepler and K2 missions, enabling us to apply Siamese and ResNet networks to our preprocessed data. This has led to the improvisation of the data and algorithms with new approaches as follows in the following sections.
     \item By utilising the back-propagation algorithm (improving performance and reducing the risk of over-fitting) and feature extraction methods, Siamese and ResNet networks can be effective in processing fewer exoplanet data. In situations when there is a limited quantity of data available and the objective includes similarity matching, verification, or identification, Siamese networks may generally be quite successful. While Siamese networks offer several characteristics that can make them particularly useful in some applications, CNN and ResNet can also be successful in similar situations.
     \item Instead of using traditional classification tasks where each input is allocated to a specific class, Siamese networks are often utilised for tasks that entail determining how similar or different two input pairs are to one another. Confusion matrices are therefore rarely used to assess the effectiveness of Siamese networks. This is our main purpose to deal with Siamese so that we get greater accuracy for each class of the planets.
     \item The usage of a three-class (Candidate, Confirmed, False-positive) classification instead of two classes (Planet, Not a planet) helps us to classify the potential candidates from False-positives with some greater accuracy. Instead of treating each light curve as a separate light curve file, we stitched the light curve files of all the Kepler quarters into a single light curve file. The single stitched light curve possesses more information than the unstitched multiple light curve files and also the stitched light curves improve the quality of our input data.

 \end{itemize}

\section{Data}
\paragraph*{} \quad The time-series data can be of three types, i.e. Periodic, transient, and stochastic. The periodic astronomical objects are planets, comets, pulsars, solar cycles, and binary stars. The transient astronomical objects are novae, supernovas, stellar activity, and gamma-ray bursts. The stochastic astronomical objects are accreting systems such as neutron stars and black hole jets. Machine Learning (ML) plays an important role in the classification of this kind of transit-shaped signal (\cite{Thompson_2015}). In this paper, the exoplanet Light Curve (LC) data from the Kepler space telescope is periodic and transient data. The light curve product is the table containing the normalised flux at each observation time. The LC's being referred to are the PDCSAP flux LC's, which have been pre-processed to remove unwanted data and noise. These PDCSAP flux LC's are considered to be a cleaner and more refined version of the original Simple Aperture Photometry (SAP) flux LC's, which may contain more artifacts and errors  (\cite{hinners2018machine}). The Kepler Input Catalog (KIC) ID number and EPIC ID for K2 data have been used to download and organise the light curves. Additionally, the light curves can be examined by plotting the time series data. The PDCSAP flux is the plot between the Flux ($e^{-1}s^{-1}$) and Time (BKJD days). Next to preprocessing, the final light curve plots are between the Normalised flux ($e^{-1}s^{-1}$) and Phase (JD). These light curves are also called Threshold Crossing Events (TCEs). TCEs are the event (similar to the solar eclipse) of a planet passing in front of the parent star, which will give a transit dip in the flux of the star concerning time (\cite{twicken2018kepler}). Transit events can be studied using light curves (\cite{2018ascl.soft12013L}). This event can be recorded and downloaded through a few lines of code in Comma-Separated Files (CSV) formats. The CSV files contain the following columns "Time", "Flux", and "Flux error". This telescopic data can be divided into two types: Kepler and K2 data. The data is taken from the same Kepler telescope but with a different field of view as mentioned earlier in the introduction.

\paragraph*{} \quad The data used by the Astronet is around 1,50,000 light curves (unstitched light curves) of the Kepler mission. These light curves are extracted from the raw Flexible Image Transport System (FITS) file and further preprocessed as mentioned in (\cite{Shallue_2018}) for the Astronet-K2 algorithm. The light curves have been classified as three varieties of views Global(G), Local(L), and combined Global-Local (GL). For the Global view, the CSV file has a fixed length of 2001 bins, and for the local view, it is 201 bins (\cite{2018ascl.soft12013L}). Similarly, the GL view combines global and local views for each light curve. These three views (G, L, GL) will help us briefly understand different models of ML algorithms’ performance.

\paragraph*{} \quad The data made publicly available by NASA (\cite{article}) contains many useful features that need to be extracted and interpolated. We have derived the data for both Kepler and K2 from the labelled candidate catalogue for Cumulative Kepler Object of Interest (KOI), which is hosted at the NASA Exoplanet Archive (\cite{article}). The three classes are named as CANDIDATE, CONFIRMED, and FALSE POSITIVE for different KepID and EPIC\_ID.

\section{Preprocessing}
\paragraph*{} \quad The preprocessing methods used by \cite{Shallue_2018}, \cite{dattilo2019identifying} are the same and are as follows. The flux of both the unstitched light curves has been normalised to have a median of 0 and minimum value of -1 so that all the TCEs have fixed transit depth. After normalisation, the out-liners have removed utilizing a sigma clipping technique and masking the transit point to avoid self-subtraction of the planet signal. Later the flattened light curve has been folded at a particular period (\textbf{tce\_period}). The light curves are flattened using the Savitzky-Golay filter (\cite{ilin2021altaipony}) to remove the long-term trends. With the support of planet orbital period (\textbf{tce\_period}) values, Transit Epoch (\textbf{tce\_time0bkjd}) and other values from the main cumulative CSV file, the light curves have been successfully folded. For K2 data, the planet orbital period (\textbf{pl\_orbper}) column is used for folding the light curves. Also, for K2 data, the Self Flat Fielding method (SFF) has been used (\cite{hedges2019four}) to remove the spacecraft motion noise. Using the SFF corrector, the signal-to-noise ratio in our light curve data has been improved. The global and local data can be created and plotted by binning the light curves. 

\paragraph*{} \quad We have used the same preprocessing methods as above, except we used the lightkurve python library (\cite{2018ascl.soft12013L}) (\cite{barentsen2020lightkurve}). By using this lightkurve package, we downloaded, preprocessed and saved the data files in CSV format to our local computer. The flux data is our input data rather than the FITS file extraction of the light curve data. As a result, a total number of 7,972 (for Kepler) and 1,199 (for K2) light curves have been downloaded. The Linear interpolation method has been used for some CSV files to evaluate the missing NaN values. These final normalised flux plots are the input data to different neural networks.

\section{Machine Learning Algorithms}
Over the past few years, several Machine Learning (ML) approaches have been proposed to speed up the exoplanet discovery process, including Decision Trees, Random Forest Classifiers (RFCs) and Convolutional Neural Networks (CNNs) (\cite{Shallue_2018}). We are using CNN-based models considering different aspects as follows:
 \begin{itemize}
 
     \item A model using the original dataset but in a lesser amount without losing its actual features.
     \item Using the feature selection and dimensionality reduction method to clean our data so that only the filtered data is preprocessed further with relevant features required in detection and classification phenomenon. 

     % (\textbf{(Feature Selection refers to the technique of decreasing the number of input variables used in your model by eliminating irrelevant or noisy data and retaining only the pertinent information. This process involves the automatic selection of relevant features for your machine learning model, which is based on the nature of the problem that you are attempting to address.)}
     
     \item In addition to the CNN model, we also employed ResNet on the same dataset to verify our findings. This approach could aid in comparing the accuracies of both models that were designed for exoplanet classification.
     \item Siamese networks compare two inputs to produce a similarity score using a neural network that has the same architecture and weights for each input. They were first introduced in the 1990s to recognize handwritten signatures and have since been used in image and text classification, face recognition, and similarity-based search. Siamese networks require fewer data than traditional supervised learning models because they compare pairs of inputs instead of categorizing them, making them useful for tasks with limited labelled data. Incorporating Siamese networks into larger models has led to state-of-the-art results on benchmark datasets. Siamese networks are a flexible and powerful tool for machine learning tasks, particularly in situations where data is scarce or costly.
     
 \end{itemize}

 \paragraph*{} \quad ML is classified as supervised, unsupervised, and reinforcement learning. For our data, we will be using supervised learning. The classification algorithm is a supervised learning technique used to identify the category of new observations based on the training data. In the classification process, algorithms learn from the given dataset and then classify new observations into several classes or groups, such as exoplanets or not an exoplanet. The main goal of the classification algorithm is to identify the category of a given dataset, and these algorithms are mainly used to predict the output for the particular input data. 
 \paragraph*{} \quad We can classify our data into different classes, and these classes have features that are similar and dissimilar to each other (\cite{Malik}). The classifier can be classified into two classes: binary and multi-class. i) Binary class classifier is a classification problem in which only two possible classes will be considered for classification problem. Likewise, in our work, we consider two classes mainly termed planet and non-planet. ii) Multi-class classifier is the classification method with more than two possible outcomes. In our work, we consider both Binary class classifiers and Multi-class classifiers. For the Multi-class classifier, we mainly classify three classes into Candidate, Confirmed, and False positive.

\paragraph*{} \quad We have chosen Convolutional Neural Network (CNN), Residual Network (ResNet), and Siamese Neural (SNN) Network to analyse the output from the space telescope data using a light curve. We discuss these networks, and their individual and Combined results in our work. The performance of the model has been evaluated using the following parameters as Accuracy, Precision, Recall and F1 score (\cite{priyadarshini2021convolutional}.  
 
\subsection{Convolutional Neural Network (CNN)}
A typical form of neural network used in image processing tasks including object identification, picture classification, and segmentation is the convolutional neural network (CNN). A high-level API for creating CNN's is provided by the TensorFlow library, which enables users to quickly specify the network's architecture and train it using data. This neural network uses filters to extract features from original images. It does this operation so that the information of the pixels is retained. It is different from classical image recognition, where we define the image features by ourselves. But in CNN, we deal with the raw image data in the pixel form (where each image is represented in the form of an array of pixel values), train the model, and extract the features automatically for better classification and detection (\cite{visser2021one}). We can illustrate the working of CNN with an example. Suppose there is an image of a bird, and we want to find out whether it is an image of a bird or something else.
\paragraph*{} \quad Firstly, the image pixels are fed in the form of arrays to the input layer of the neural network. The hidden layers extract the features for performing different calculations and manipulations. Various hidden layers like the convolution layer, the Re-LU layer, and the Pooling layer perform feature extraction from the image. Finally, the last layer, called the fully connected layer, identifies the object in the image as an output.\textcolor{blue}{} Filters are applied repeatedly to improve the efficiency of training (\cite{patel2020comprehensive}. Some previous research has shown accuracy levels above 90\%, and CNN's have demonstrated promising outcomes in applications like exoplanet discovery and categorization of light curves. Layers in CNN are in the following order: first comes the input layer, Convo layer(Convo+Re-LU), Pooling layer, Fully Connected (FC) layer, Softmax/Logistic layer, and output layer. We will discuss these different layers in detail in the below section.

% \section*{Layers in a Convolutional Neural Network}
% \textbf{Layers in a Convolutional Neural Network}
\paragraph*{} \quad A convolution neural network has more than one hidden layer that helps in extracting information from an input image. The four crucial hidden layers in CNN are

\begin{itemize}
    \item \textbf{Convolution Layer: } It is the first process in extracting features from an image. It has several filters that perform the convolution process. Every single image is considered a matrix of pixel values.
    \item \textbf{Re-LU Layer: } It stands for the rectified linear unit. Re-LU performs an element-wise operation and sets all the negative values to 0 pixels. The generated output in this layer is a corrected feature map.
    \item \textbf{Pooling Layer: } The operation reduces the dimensionality of the feature map. The layer uses various filters to identify different parts of the image. Flattening in this network converts all the 2D arrays into a single long continuous linear vector.
    \item \textbf{Fully connected Layer: } The flattened matrix is fed to the fully connected layer as an input to classify the image to get the final output.
\end{itemize}

\paragraph*{} \quad CNN is used to understand the convolution operation, to understand the pooling operation, remember the stride, filter, padding, etc., and build a CNN for multi-class classification in images. CNN is mainly used for 2D image classification problems. And also, have 1D CNN to handle other types of data. Its architecture consists of a stack of 17 distinct layers that transform the input volume into an output volume through a differential function that is important because it allows us to back-propagate the model's error when training to develop the weights. The first layers are a mixture of convolutional, max pooling, and batch normalisation layers used to extract different patterns in the time series. Dropout layers are applied to prevent the model from over-fitting to the training data. We have used 1D flux data from the CSV files as input for the CNN algorithm and all others. 

\subsubsection{CNN using Astronet-K2}
\begin{figure}[htp]
    \centering
    \includegraphics[height=22cm, width=9cm]{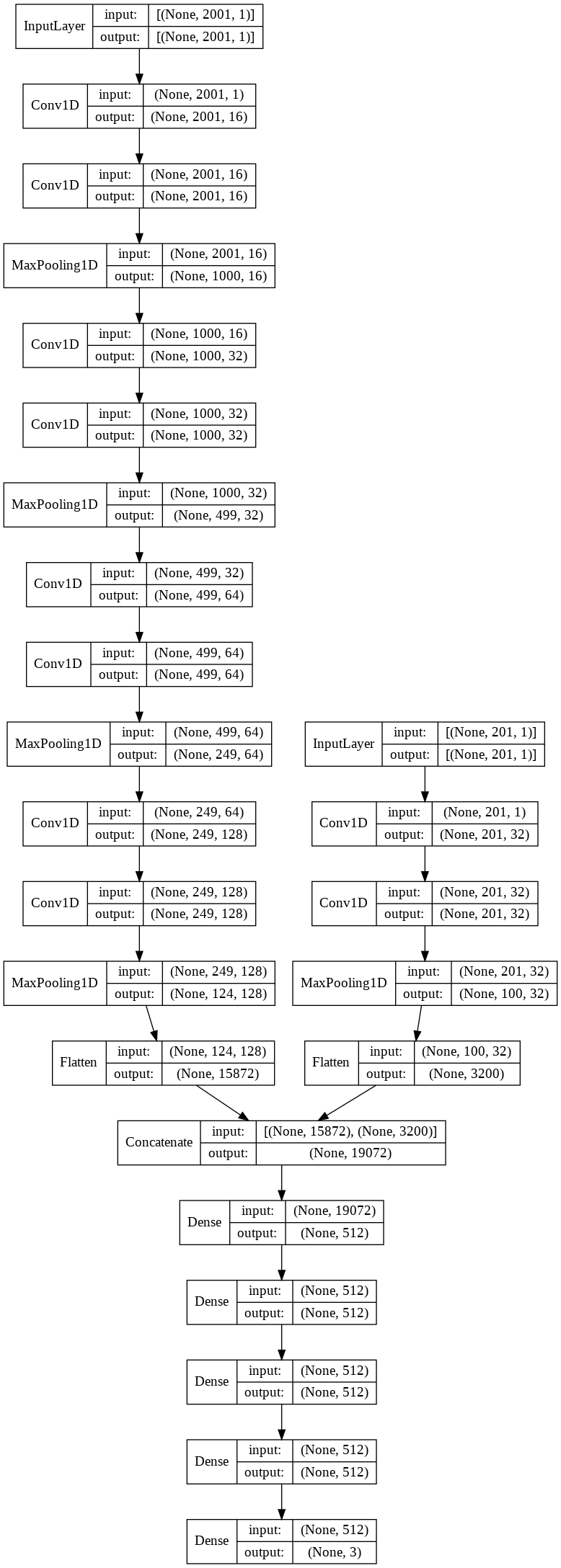}
    \caption{CNN architecture for GL data for three-class classification using Keras plot model function.}
    \label{fig:CNN_architecture}
\end{figure}

Better accuracy, precision, recall, and f1 score results have been achieved by modifying the Astronet-K2 algorithm for our data. The results have been tabulated in Table \ref{tab:CNN_tf_2class_astronet} for 2-Class classification. The astronet-K2 algorithm is created on the background of Tensor-flow version 1. So, for this algorithm, the CSV files of our data had to be converted into Tensor-flow (TF) records with eight train files, one validation file, and 1 test file. To record is a simple format file used to store data as a sequence of binary records. This conversion of data into TFRecord has a few advantages, such as it can take up less space than the original data and can also be partitioned into multiple files. With assistance from the training and evaluating bazel code \cite{Shallue_2018}, the training and testing have been achieved with the model being saved to our specified model directory. This model uses only Global-Local (GL) view data with 625 epochs. As shown in figure \ref{fig:CNN_architecture}, the model consists of various fully connected and convolutional neural networks as represented in the (\cite{dattilo2019identifying}) paper. This model has been trained and tested for various train-test split ratios for our paper to compare the data preprocessing quality and to help us improve further a better model for the classification of exoplanets. Our results show good accuracy following the previously published paper (\cite{dattilo2019identifying}) using the astronet-K2 code. 

\begin{longtable}[t]{|c|c|c|c|c|c|c|}
\caption{CNN results for a Two-class classification.}
\label{tab:CNN_tf_2class_astronet}\\
\hline
 Mission and TTR & Data Types & Accuracy & Precision & Recall & F1 Score\\
 \hline
 \endfirsthead

 \hline
 \endhead

 \hline
 \endfoot

 \hline
 \endlastfoot
Kepler (90:10) & GL & 0.89 %0.8861893 
& 0.91 & 0.92 & 0.92 \\
% Kepler (80:20) & GL & 0.8804348 & 0.58 & 0.47 & 0.52 \\
% Kepler (70:30) & GL & 0.8695652 & 0.51 & 0.49 & 0.50 \\
Kepler (60:40) & GL &  0.87 % 0.870844 
& 0.93 & 0.89 & 0.91 \\
Combined (90:10) & GL &  0.84 %0.8363821 
& 0.86 & 0.89 & 0.87 \\
Combined (80:20) & GL &  0.82 %0.817997 
& 0.91 & 0.83 & 0.87 \\
Combined (70:30) & GL &  0.82 %0.8214165 
& 0.93 & 0.83 & 0.88 \\
% Combined (60:40) & GL & 0.8332486 & 0.37 & 0.45 & 0.41 \\

\end{longtable}
  \begin{tablenotes}
    \small
    \item \textbf{Note --}$^\textbf{}${This table represents the Kepler and combined data for various train test split ratios for two-class classification using the CNN algorithm (TensorFlow version 1). This includes the light curve data of the Global-Local(GL) view.}

  \end{tablenotes}

\subsubsection{CNN using Keras}
To further contribute to the CNN algorithm, the Astronet-K2 model has been created in terms of Keras background as it is easy to handle and modify according to our convenience (\cite{gulli2017deep}). Our 1D flux data is separated and identified by the convolution layer, which separates and identifies the numerous extraction characteristics for analysis. In our scenario, two convolution layers are followed by a Max-pooling layer. After flattening the convolutional layers, the neuron is connected to four dense layers of 512 neurons were used. In the case of GL, this is the step at which we integrate the G and L data for classification. The last dense layer consists of 2 or 3 neurons depending on our class and also uses the output from these four Dense layers to predict the exoplanet class using the features extracted in the preceding procedure. The data is categorised into two (Planet and not a planet) classes and three (Candidate, Confirmed, False-positive) classes using Keras, for all three views (GL, G, L). The training and testing process has been carried out for various train-test split ratios for 300 epochs. The results are shown in table \ref{tab:CNN_keras_2class} (2-Class) and \ref{tab:CNN_keras_3class} (3-Class) for both Kepler and Combined data with various accuracy, precision, recall and F1 score values. In the case of 3-Class classification, the values of Candidate, Confirmed, and False-positive are represented by three rows of precision, recall, and F1 score values in table \ref{tab:CNN_keras_3class}. It is the same for ResNet 3-class classification table as shown below.

\begin{longtable}[c]{|c|c|c|c|c|c|c|}
\caption{CNN results for a Two-class classification.}
\label{tab:CNN_keras_2class}\\
\hline
 Mission and TTR & Data Types & Accuracy & Precision & Recall & F1 Score\\
 \hline
 \endfirsthead

 \hline
 \endhead

 \hline
 \endfoot

 \hline
 \endlastfoot

Kepler (90:10) & GL & 0.84 %0.8446 
& 0.76 & 0.69 & 0.72 \\
& G & 0.85 %0.8496  
& 0.72 & 0.72 & 0.72 \\
& L& 0.84 %0.8446  
& 0.73 & 0.70 & 0.71 \\
Kepler (80:20) & GL & 0.83 %0.8326 
& 0.71 & 0.69 & 0.70 \\
& G & 0.83 %0.8320  
& 0.69 & 0.69 & 0.69 \\
& L & 0.84 %0.8433  
& 0.72 & 0.71 & 0.72 \\
Kepler (70:30) & GL & 0.82 %0.8232 
&  0.73 & 0.65 & 0.69 \\
& G & 0.84 %0.8403 
& 0.74 & 0.69 & 0.72 \\
& L & 0.83 %0.8328 
& 0.70 & 0.69 & 0.69 \\
Kepler (60:40) & GL & 0.84 %0.8394 
& 0.72 & 0.70 & 0.71 \\
& G & 0.82 %0.8213 
& 0.70 & 0.67 & 0.68 \\
& L & 0.85 %0.8460 
& 0.75 & 0.71 & 0.73 \\
Combined (90:10) & GL & 0.82 %0.8246 
& 0.64 & 0.70 & 0.67 \\
& G & 0.82 %0.8028 
& 0.56 & 0.67 & 0.61 \\
& L & 0.82 %0.8214 
& 0.63 & 0.69 & 0.66 \\
Combined (80:20) & GL & 0.82 %0.8202 
& 0.66 & 0.68 & 0.67 \\
& G & 0.79 %0.7913 
& 0.62 & 0.62 & 0.62 \\
& L & 0.83 %0.8278 
& 0.63 & 0.71 & 0.67 \\
Combined (70:30) & GL & 0.81 %0.8081 
& 0.66 & 0.67 & 0.66 \\
& G & 0.80 %0.8005 
& 0.57 & 0.68 & 0.62 \\
& L & 0.83 %0.8292 
& 0.66 & 0.72 & 0.69 \\
Combined (60:40) & GL & 0.83 %0.8261 
& 0.68 & 0.70 & 0.69 \\
& G & 0.80 %0.8043 
& 0.62 & 0.67 & 0.64 \\
& L & 0.82 %0.8220 
& 0.64 & 0.71 & 0.67 \\

\end{longtable}
\begin{tablenotes}
      \small
       \item \textbf{Note --}$^\textbf{}$ {This table represents the Kepler and combined data for various train test split ratios for two-class classification using the CNN algorithm (Keras version). This includes the Global(G), Local(L) and Global Local(GL) views.}
\end{tablenotes}

\begin{longtable}[c]{|c|c|c|c|c|c|c|}
\caption{CNN results for a Three-class classification.}
\label{tab:CNN_keras_3class}\\
\hline
 Mission and TTR & Data Types & Accuracy & Precision & Recall & F1 Score\\
 \hline
 \endfirsthead

 \hline
 \endhead

 \hline
 \endfoot
 
Kepler (90:10) %& GL & 0.6842 &  0.52 & 0.46 & 0.49 \\
% & & & 0.63 & 0.78 & 0.70 \\
% & & & 0.81 & 0.75 & 0.78 \\
& G & 0.69 %0.6930 
&  0.53 & 0.48 & 0.50 \\
& & & 0.67 & 0.73 & 0.70 \\
& & & 0.79 & 0.79 & 0.79 \\
% & L & 0.6291 &  0.45 & 0.37 & 0.41 \\
% & & & 0.60 & 0.73 & 0.66 \\
% & & & 0.74 & 0.71 & 0.73 \\
% Kepler (80:20) & GL & 0.6771 &   0.50 & 0.43 & 0.46 \\
% & & & 0.69 & 0.72 & 0.70 \\
% & & & 0.75 & 0.79 & 0.77 \\
% & G & 0.6784 &  0.51 & 0.46 & 0.48 \\
% & & & 0.66 & 0.72 & 0.69 \\
% & & & 0.77 & 0.78 & 0.77 \\
% & L & 0.6803 &  0.51 & 0.40 & 0.44 \\
% & & & 0.71 & 0.75 & 0.73 \\
% & & & 0.74 & 0.80 & 0.77 \\
Kepler (70:30) & GL & 0.69 %0.6881 
&   0.53 & 0.50 & 0.51 \\
& & & 0.66 & 0.73 & 0.69 \\
& & & 0.79 & 0.78 & 0.78 \\
% & G & 0.6760 & 0.51 & 0.45 & 0.48 \\
% & & & 0.66 & 0.70 & 0.68 \\
% & & & 0.77 & 0.79 & 0.78 \\
% & L & 0.6593 & 0.50 & 0.39 & 0.44 \\
% & & & 0.65 & 0.74 & 0.69 \\
% & & & 0.73 & 0.77 & 0.75 \\
% Kepler (60:40) & GL & 0.6886 &   0.51 & 0.43 & 0.47 \\
% & & & 0.66 & 0.78 & 0.72 \\
% & & & 0.79 & 0.78 & 0.78 \\
% & G & 0.6729 & 0.51 & 0.38 & 0.44 \\
% & & & 0.63 & 0.75 & 0.68 \\
% & & & 0.77 & 0.79 & 0.78 \\
% & L & 0.6676 & 0.49 & 0.43 & 0.46 \\
% & & & 0.66 & 0.73 & 0.69 \\
% & & & 0.76 & 0.76 & 0.76 \\
Combined (90:10) %& GL &  0.6275 &   0.45 & 0.49 & 0.47 \\
% & & & 0.67 & 0.65 & 0.66 \\
% & & & 0.73 & 0.70 & 0.71 \\
& G & 0.63 %0.6285  
&  0.50 & 0.51 & 0.50 \\
& & & 0.58 & 0.67 & 0.62 \\
& & & 0.76 & 0.68 & 0.72 \\
& L & 0.63 %0.6296    
&  0.49 & 0.52 & 0.51 \\
& & & 0.61 & 0.61 & 0.61 \\
& & & 0.74 & 0.71 & 0.72 \\
Combined (80:20) & GL & 0.66 %0.6583  
&   0.51 & 0.56 & 0.53 \\
& & & 0.66 & 0.68 & 0.67 \\
& & & 0.78 & 0.71 & 0.75 \\
% & G & 0.6343  &  0.52 & 0.45 & 0.49 \\
% & & & 0.56 & 0.67 & 0.61 \\
% & & & 0.77 & 0.74 & 0.75 \\
& L & 0.65 %0.6452   
&  0.51 & 0.51 & 0.51 \\
& & & 0.64 & 0.71 & 0.67 \\
& & & 0.75 & 0.70 & 0.72 \\
Combined (70:30) & GL & 0.67 %0.6701  
&   0.54 & 0.52 & 0.53 \\
& & & 0.64 & 0.69 & 0.66 \\
& & & 0.78 & 0.76 & 0.77 \\
& G & 0.64 %0.6352   
&  0.50 & 0.52 & 0.51 \\
& & & 0.62 & 0.60 & 0.61 \\
& & & 0.75 & 0.74 & 0.74 \\
% & L & 0.6344  &  0.52 & 0.45 & 0.48 \\
% & & & 0.60 & 0.71 & 0.65 \\
% & & & 0.74 & 0.71 & 0.73 \\
Combined (60:40) & GL & 0.66 %0.6615  
&   0.54 & 0.49 & 0.51 \\
& & & 0.68 & 0.66 & 0.67 \\
& & & 0.72 & 0.79 & 0.76 \\
& G & 0.66 %0.6555  
&  0.68 & 0.49 & 0.57 \\
& & & 0.61 & 0.66 & 0.63 \\
& & & 0.75 & 0.88 & 0.81 \\
& L & 0.63 %0.6334  
&  0.48 & 0.55 & 0.51 \\
& & & 0.66 & 0.60 & 0.63 \\
& & & 0.75 & 0.72 & 0.73 \\
\end{longtable}
\begin{tablenotes}
  \small
  \item \textbf{Note --}$^\textbf{}$ {This table represents the Kepler and combined data for train test split ratio of 70:30 for three-class classification using the CNN algorithm (Keras version). This includes the light curve data of the Global(G), Local(L) and Global Local(GL) views.}
\end{tablenotes}
 
\subsection{Residual Network (ResNet)}
\begin{figure}[htp]
    \centering
    \includegraphics[height=22cm, width=15cm]{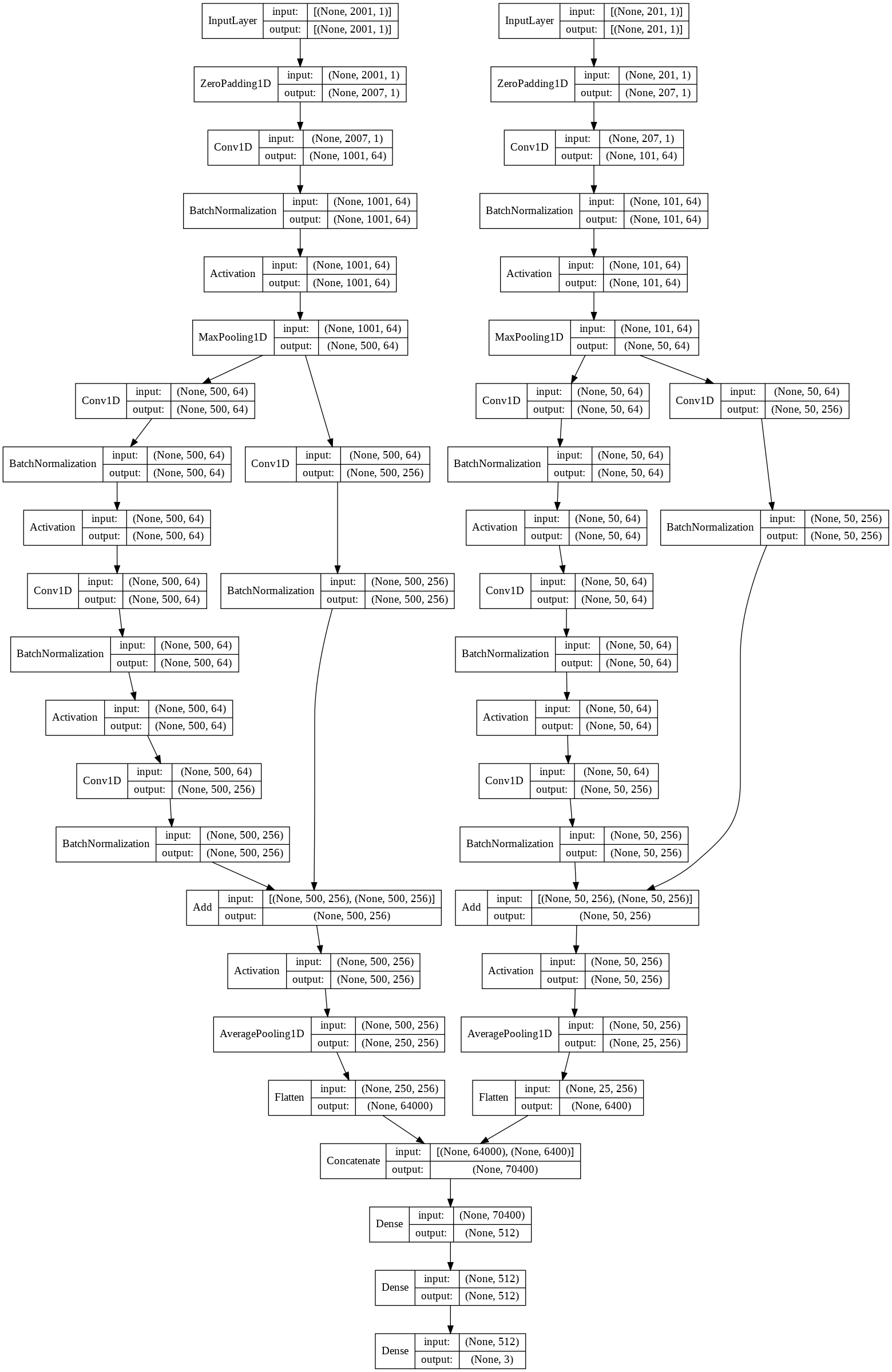}
    \caption{ResNet architecture for GL data for three-class classification using Keras plot model function.}
    \label{fig:ResNet_architecture}
\end{figure}

Adding extra layers to a suitably deep neural network causes accuracy to reach saturation and subsequently decrease. To solve this problem of the vanishing gradient or degradation problem, Residual-Network (ResNet) was proposed by researcher Kaiming He (\cite{he2016identity}) at the ILSVRC 2015 competition. It is a deep convolutional network comprised of a novel approach pathway called skip connection. The basic idea is to skip blocks of convolutional layers by using short connections. These connections provide an alternate pathway for data and gradients to flow and thus making the training possible (\cite{he2016identity}. Deep neural network architecture ResNet-50 has been extensively utilised in picture identification applications, including the categorization of exoplanets. A broad outline of the procedures involved in classifying exoplanets using ResNet-50: Prior to anything else, we prepared our data (dataset of exoplanet images). Preprocessing the data (e.g., normalising, resizing) and dividing the dataset into training, validation, and test sets may be necessary to do this. Each ResNet layer is two (ResNet-18, 34) or three layers deep (ResNet-50, 101, 152).
\paragraph*{} \quad With the specific code, we can load the ResNet-50 model from the TensorFlow/Keras package. Then Customize the top layers and the last and after that, by training the ResNet-50 model on our exoplanet dataset, we may fine-tune it. The ResNet-50 model's layers may be frozen up to a certain point, and only the newly added layers need to be trained on our dataset. By doing so, you may expedite the training process and avoid over-fitting. Lastly, we can assess the ResNet-50 model's performance on our test data using measures like recall, accuracy, and precision. We have employed ResNet-50 with Re-LU, and batch normalisation. Batch Normalisation enables us to employ much larger learning rates while also being less cautious with initialisation. It also functions as a regulariser, obviating the necessity for Dropout in some circumstances (\cite{ioffe2015batch}). The first 1x1 convolution layer with stride 2 reduces the dimension for the feature calculation using the 3x3 bottleneck layer. Again using the last 1x1 convolution layer, the dimension has been increased back to the original version. This method of using 1x1 filters for reducing and increasing the dimensions was first in the GoogLeNet model by \cite{szegedy2014going}. We have used some simple identity blocks to change the depth, not the dimension of the depth. But the residual blocks change the input dimension and the skip connection. By combining all these blocks, we have created our ResNet-50 model. This model has completed 300 epochs for all data views (GL, G, and L), and its sample architecture is given in Figure \ref{fig:ResNet_architecture}. Unlike CNN, this algorithm is executed for a single TTR ratio of 70:30.

\begin{longtable}[c]{|c|c|c|c|c|c|}
  \caption{ResNet results for Two-class classification.}
  \label{tab:ResNet_2class}\\
  \hline
  \textbf{Mission} & \textbf{Data Types} & \textbf{Accuracy} & \textbf{Precision} & \textbf{Recall} & \textbf{F1 Score} \\
  \hline
  \endfirsthead 

  \hline
  \endhead

  \hline
  \endfoot

  Kepler & GL & 0.84 & 0.45 & 0.90 & 0.60 \\
         & G  & 0.80 & 0.56 & 0.65 & 0.60 \\
         & L  & 0.84 & 0.59 & 0.73 & 0.65 \\
  % \hline
  Combined & GL & 0.86 & 0.51 & 0.82 & 0.63 \\
           & G  & 0.78 & 0.48 & 0.64 & 0.55 \\
           & L  & 0.82 & 0.61 & 0.71 & 0.65 \\
  % \hline
\end{longtable}
    \begin{tablenotes}
          \small
          \item \textbf{Note --}$^\textbf{}$ {This table represents the Kepler and combined data for a single train test split ratio of 70:30 for two-class classification using the ResNet algorithm (Keras Background). This includes the light curve data of the Global(G), Local(L) and Global Local(GL) views.}
    \end{tablenotes}

\begin{longtable}[c]{|c|c|c|c|c|c|c|}
\caption{ResNet results for Three-class classification.}
\label{tab:ResNet_3class}\\
\hline
Mission & Data Types & Accuracy & Precision & Recall & F1 Score\\
\hline
\endfirsthead
\hline
\endhead
\hline
\endfoot
% Kepler  & GL & 0.7337 &   0.47 & 0.50 & 0.48 \\
%         &    &        &   0.86 & 0.46 & 0.60 \\
%         &    &        &   0.71 & 0.87 & 0.78 \\
%         & G  & 0.6501 &   0.57 & 0.37 & 0.45 \\
%         &    &        &   0.56 & 0.77 & 0.65 \\
%         &    &        &   0.75 & 0.74 & 0.75 \\
%         & L  & 0.6237 &   0.42 & 0.38 & 0.40 \\
%         &    &        &   0.66 & 0.62 & 0.64 \\
%         &    &        &   0.71 & 0.76 & 0.74 \\
Combined & GL & 0.76 %0.7569  
        & 0.59 & 0.68 & 0.64 \\
        &    &      & 0.80 & 0.65 & 0.72 \\
        &    &      & 0.80 & 0.81 & 0.81 \\
        & G  & 0.63 %0.6279  
        & 0.52 & 0.54 & 0.53 \\
        &    &      & 0.72 & 0.57 & 0.64 \\
        &    &      & 0.71 & 0.80 & 0.75 \\
        & L  & 0.63 %0.6312  
        & 0.48 & 0.46 & 0.47 \\
        &    &      & 0.65 & 0.69 & 0.67 \\
        &    &      & 0.72 & 0.72 & 0.72 \\
\end{longtable}
\begin{tablenotes}
  \small
  \item \textbf{Note --}\textsuperscript{} {This table represents only the combined data for a single train test split ratio of 70:30 for three-class classification using the ResNet algorithm. This includes the light curve data of the Global (G), Local (L) and Global Local (GL) views.}
\end{tablenotes}
\subsection{Siamese Network}
CNNs are indeed the best-performing methods for image classification tasks, however, the major constraint of utilising this approach in such image-related tasks is that it requires a huge amount of labelled data. In the recent Machine - learning period, neural networks are close to perfect at almost every job; unfortunately, these neural networks required more data to perform well. In many real-world applications, gathering this much data is difficult and impractical, particularly in the field of observational astronomy, but here comes one-shot learning that requires just one training example for each class. (\cite{koch2015siamese}. Similarly, for some situations, such as face recognition and signature verification, we cannot always rely on more data; to respond to these challenges, we have created a new form of neural network architecture known as Siamese Networks. The Siamese network is a one-shot classification model that can predict with only one training example.  One-shot learning consists of two symmetrical networks called the Siamese neural networks. Siamese networks are neural networks containing two identical neural networks, each taking one of the two input images and sharing common parameters and weights. Then feed the last layer of the two networks to the equivalent loss function, which calculates the similarity between the two images. It has a unique structure to naturally rank similarity between inputs (\cite{chicco2021siamese}. It is more robust to imbalanced data as it requires less information. It may be applied to a dataset with relatively few samples of some classes. Siamese networks have grown in popularity in recent years due to their ability to learn from relatively less data. The modification of parameters is synchronized across both sub-networks. It is utilised to determine the similarity of inputs by comparing feature vectors, and thus these networks are utilised in a wide range of applications.

\begin{figure}[htp]
  \centering
  \includegraphics[width=0.8\textwidth]{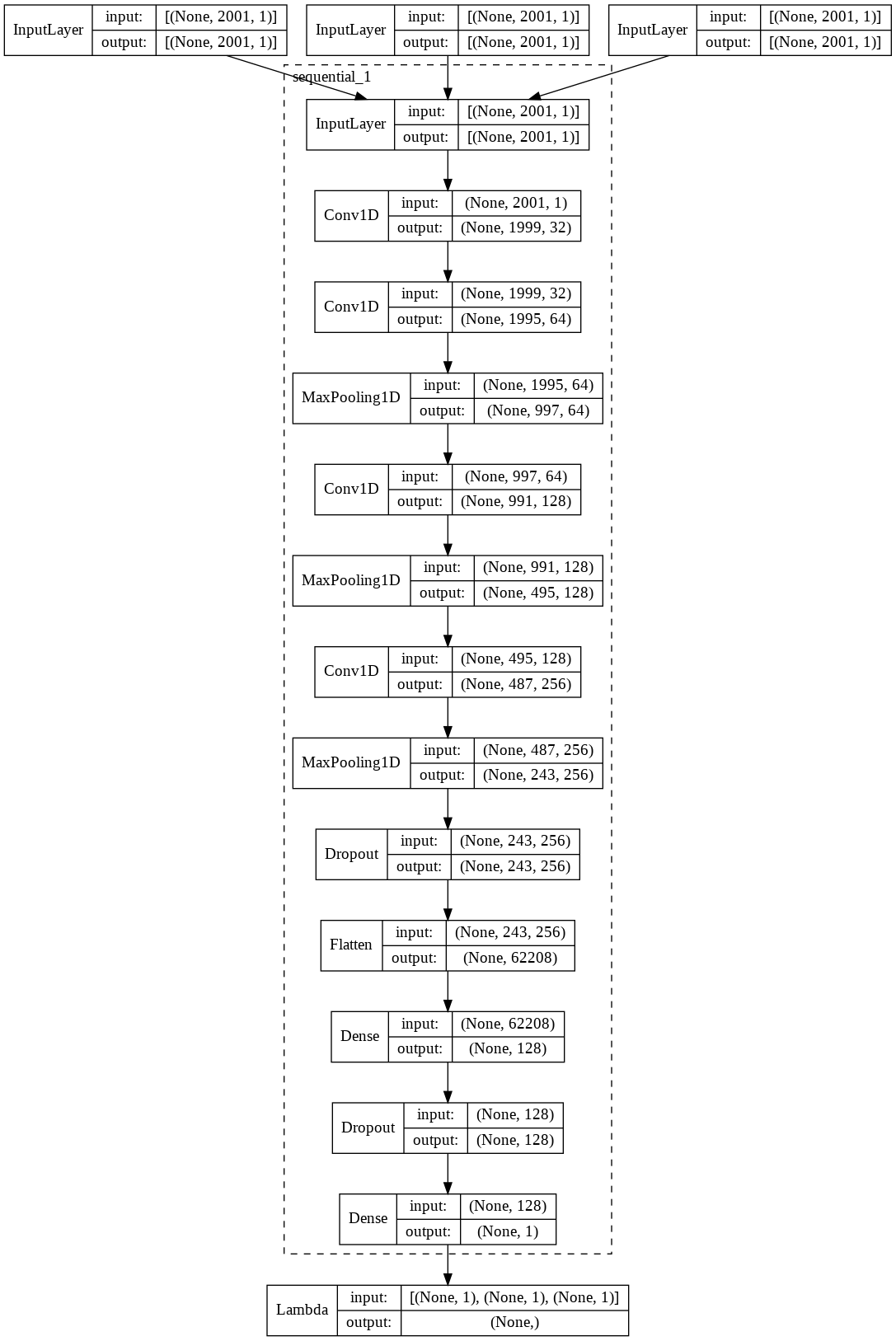}
  \caption{Siamese architecture for G data for three-class classification using Keras' plot model function.}
  \label{fig:Siamese_architecture}
\end{figure}

\paragraph*{} \quad A Siamese network may be used in the context of exoplanet classification to discover similarities between pairs of light curves, which can be helpful in spotting planetary transits. Two inputs (in this example, two light curves) are passed through identical neural networks (or "branches") that share weights in a Siamese network. The two branches' outputs are then contrasted using a distance measure, such as cosine similarity or Euclidean distance. The similarity between the two inputs is represented by a single number that the distance metric generates. One method for classifying exoplanets makes use of a Siamese network that has been trained to recognise similarities between sets of light curves, one of which has an exoplanet transit and the other of which does not. The network is taught to decrease the distance between pairs of light curves that contain exoplanet transits and maximise the distance between pairs of light curves that do not during training by presenting it with pairs of light curves and their associated labels. Using a set of reference light curves as a reference, the network may be taught to categorise new light curves. The new light curve is then assigned to the class (with or without an exoplanet transit) that has the reference light curves that are most similar to it by the network, which compares its similarity to each of the reference light curves. Siamese networks can be helpful in classifying exoplanets because they can identify intricate patterns in light curves that can be hidden by more conventional classification techniques. Siamese networks are trained on minimal datasets and become more resistant to changes in light curve shape by learning the similarity between pairs of light curves. In order to increase classification accuracy, Siamese networks can be used in combination with other classification techniques like random forests or support vector machines. Generally, the Siamese network performs a binary classification operation on the output and classifies whether the input belongs to the same class or not and different loss functions are used during their training. In many applications, two identical sub-nets are used to process two inputs, and another module will take its output and produce the final output.
 \paragraph*{} \quad Triplet loss and Contrastive loss are the two basic loss functions used in training Siamese networks. For this classification, We have used the Triplet loss function in our network to generate a generator of triplets for training or testing. Choosing a triplet (anchor, positive, negative) data such that anchor and positive have the same label while the anchor and negative have different labels. The accuracy for this classification is computed with a fixed threshold on distances. After computing loss, the model is created with three embedding models and minimizes the loss between their output embedding. Finally, the results have been summarized and given in the table \ref{tab:Siamese_2class} and \ref{tab:Siamese_3class} for 2-class and 3-class classification. 
The triplet loss and similarity distance are determined for both two and three-class classification with a 70:30 TTR and 150 epochs for this algorithm. The output value difference then varies between 0 and 1, with 1 indicating that the labelled data given to the input layers is the same and 0 indicating that it is not.  Figure \ref{fig:Siamese_architecture} depicts the Siamese model architecture for global data. 

\begin{longtable}[c]{|c|c|c|c|}
\caption{Siamese results for Two-class classification.
\label{tab:Siamese_2class}}\\
  \hline
 Mission and TTR & Data Types & Accuracy\\
 \hline
 \endfirsthead
 \hline
 \endlastfoot
Kepler (70:30) & G & 0.94 %0.94088 
\\
& L & 0.95 %0.94775 
\\
Combined (70:30) & G & 0.95 %0.95397 
\\
 & L & 0.93 %0.93255 
 \\
\end{longtable}
    \begin{tablenotes}
          \small
          \item \textbf{Note --}$^\textbf{}$ {This table represents the Kepler and combined data for a single train test split ratio of 70:30 for two-class classification using the Siamese algorithm (Keras Background). This includes the light curve data of the Global(G) and Local(L) views.}
    \end{tablenotes}
 
\begin{longtable}[c]{|c|c|c|c|}
\caption{Siamese results for Three-class classification.
\label{tab:Siamese_3class}}\\
  \hline
 Mission and TTR & Data Types & Accuracy \\
 \hline
 \endfirsthead
 \hline
 \endlastfoot
Kepler (70:30) & G & 0.94 %0.93797 
\\
& L & 0.93 %0.93013 
\\
Combined (70:30) & G & 0.97 %0.96661 
\\
& L & 0.96 %0.9573 
\\
\end{longtable}
    \begin{tablenotes}
          \small
          \item \textbf{Note --}$^\textbf{}$ {This table represents the Kepler and combined data for a single train test split ratio of 70:30 for three-class classification using the Siamese algorithm (Keras Background). This includes the light curve data of the Global(G) and Local(L) views.}
    \end{tablenotes}
\section{Results and discussion}
The training and testing of the raw data in a single and mixed form of Kepler and K2 data have been used for various ML algorithms. For each model mentioned in this work, the data is trained, validated, and tested in all three neural networks in batches of different epochs and different TTR. 
\begin{itemize}
    
    \item \textbf{CNN (Tensorflow version): }With the use of roughly 1,50,000 light curve data, scientists (\cite{Shallue_2018}) (\cite{dattilo2019identifying}) were able to attain 96\% in earlier research. However, using well-preprocessed and limited data, we attained 89\% accuracy for Kepler data with a 0.92 F1 Score, and 84\% accuracy with a 0.87 F1 Score for combined data (Kepler and K2 both) with a 90:10 train-test split ratio (TTR). Table \ref{tab:CNN_tf_2class_astronet} for two-class classification using TensorFlow version 1 provides a detailed accuracy, function, precision, recall, and f1 score for the CNN algorithm at different train-test split ratios for GL view. Similarly, some good accuracy and F1 Score is been observed for Kepler (60:40), Combined (80:20) and Combined (70:30).  
    
    \item \textbf{CNN (Keras version): }The overall findings in Kepler (90:10) from table \ref{tab:CNN_keras_2class} showed higher accuracy for all three views, with GL receiving a 0.72 F1 Score. TTR (60:40) has also demonstrated overall improved accuracy with combined data, achieving a 0.69 F1 Score for GL. We may deduce from the table \ref{tab:CNN_keras_3class} that the GL view classification is more accurate than the G and L views and has a higher F1 Score for Confirmed and False-Positive categorisation. It offers us optimism that, with further improvisation of the method, we could be able to achieve 3-class classification utilising GL data together, as \cite{dattilo2019identifying} did for 2-class. At the same time, the F1 Score of the Candidate and Confirmed classification is low and the code needs to be improvised further for better classification of these 3-Class classifications. The F1 score of 0.5 or above is considered reasonable for most classification tasks. However, the optimal F1 score will depend on the specific requirements of the task at hand, and it is important to consider other factors such as the consequences of false positives and false negatives, the balance between precision and recall, and the trade-off between model complexity and performance. The accuracy, precision, recall, and f1 score values for two and three class classifications using Keras for the same CNN method are shown in tables \ref{tab:CNN_keras_2class} and \ref{tab:CNN_keras_3class}. The three-class classification using Keras for Astronet-K2 is used in this article to help better comprehend the data.
    
    \item \textbf{ResNet: }In the 2-Class classification, the accuracy and F1 score for the L view exhibit higher results, as shown in table \ref{tab:ResNet_2class} for both Kepler and combined data. This algorithm for 2-Class classification has failed to produce higher accuracy and F1 score for GL data type as observed previously in the CNN algorithm. However, as shown in table \ref{tab:ResNet_3class}, GL values exhibit greater accuracy and F1 score for 3-Class classifications. The higher accuracy and low F1 Score suggest that the machine learning algorithm may not be performing optimally on the given data. The F1 score takes into account both precision and recall, which are important metrics in evaluating classification algorithms. A low F1 score can indicate that the algorithm is either not identifying all instances of a given class (low recall) or is identifying instances incorrectly (low precision). Apart from this, there are some possible issues that could cause a low F1 score such as Imbalanced classes, Incorrect feature selection, Insufficient data and Model complexity. Overall, for this algorithm, it is important to analyze the data and algorithm carefully to identify potential issues and improve the model's performance. The accuracy, precision, recall, and F1 score for two and three-class classifications are shown in tables \ref{tab:ResNet_2class} and \ref{tab:ResNet_3class}.
    
    \item \textbf{Siamese: }This neural network has managed to attain 99\% for both the data (Kepler and Combined) and the views (G, L). The accuracy, Mission, TTR and Data typers are tabulated in the tables \ref{tab:Siamese_3class} and \ref{tab:Siamese_2class}. The Siamese Network has been executed for the same train-test split ratio as the Residual network, but with two view types (G and L). For two-class classification, the accuracy ranges between 93\% to 95\%. For three-class classification, the accuracy is in the range of 93\% to 96\% for both Kepler and combined data. It gives us hope for a successful 3-Class classification of exoplanets using light curves. The low F1 score observed in ResNet can be attributed to the use of insufficient data in the algorithm. To address this issue, researchers can focus on improving and modifying the Siamese algorithm to achieve better classification results for low-data scenarios such as exoplanet data classification or other astrophysical objects. 
    
\end{itemize}
In two-class classification, False-positive and Candidate classes have been considered as a single, not a planet-class, but three classes, it is considered separately. It has been noted that the three-class classification has accuracy far less than the two-class classification. It is due to the confusion that arises between the classification of the candidate and confirmed light curves. The main aim of this paper is not to increase the accuracy or reduce the loss function; instead, to observe the variation of accuracy and the performance of the different algorithms for various train-test split ratios. It will help us learn more about the data and will help us improve our techniques of data collection and preprocessing methods. In addition, different TTR values help to mark the accuracy and F1-Score without biasing any data. These algorithms in the identification and classification of exoplanets will help our fellow scientists have a brief knowledge of the data acquired and improvise these data-collecting methods further. Although extraction features such as planet orbital period, Transit Epoch, and Transit duration might have increased accuracy, precision, recall, and f1 score, the paper aims to discover the best-automated algorithm for exoplanet classification using just light curve data. By leveraging advancements in machine learning techniques and developing more sophisticated algorithms, we can potentially unlock new insights and discoveries in the field of astronomy and astrophysics. The use of high-quality, diverse datasets along with powerful machine learning models can help us extract meaningful patterns and knowledge from the vast and complex universe that surrounds us.

\section{Future Mission and Methods}
With further recent improvements in ML techniques, we can even use 3D synthesis techniques (\cite{han2018improving}) to make a detailed study of the exoplanets apart from the classification and detection problem. The detailed analysis includes the star's distance, planet, and star ratio related to our solar system and automatically mapping the stars using different parameters other than flux from the NASA Exoplanet Archive website. With various flux measurements, The classification of the type of the star is possible for both the Kepler and K2 data using multiple machine learning algorithms (\cite{armstrong2015K2}) (\cite{davies2015oscillation}). The properties like its main-sequence star, red giant star, blue giant star or white dwarf can be classified with these algorithms. We also launched Transiting Exoplanet Survey Satellite (TESS) on April 18 2018. It is specially launched with the motive to find more Earth-like planets, and the satellite is far better than the Kepler space telescope. The area covered by the TESS mission is 400 times larger than the area covered by Kepler to search for planets within approximately 200 light-years. Stars discovered by the TESS mission are about 300-100 times brighter than those surveyed by the Kepler satellite. With this launch, we believe we can find the right candidate for an Earth-like planet so that we can call it our second home planet (\cite{stassun2018tess}). The sky is the limit when it comes to the possibilities of this technology. So far, Kepler has discovered a total of 4780 candidates and confirmed planets whereas, the confirmed planets discovered by the K2 mission so far is 463 in number, with 889 K2 candidates yet to be confirmed. There may be many exoplanets yet to be found in Kepler and K2 data. With new ideas and various techniques like machine learning, Deep Learning will help fuel celestial discoveries for many years to come. To infinity and beyond!

\section{Conclusion} 
It's incredible to think that with the help of light obtained from faraway stars, researchers may examine this light that has travelled hundreds of light-years and draw conclusions about what possible planets these stars might harbour. We can calculate the star-planet radius ratio, planet mass, and orbital period analytically using exoplanet light curve data. In astronomy, this type of time-series data is the most prevalent. The scientists' primary focus is to create a single machine-learning system that can classify all-time series objects like pulsars, supernovae, Cepheid variables, and planets transiting parent stars. Our method for detecting and classifying exoplanets using a machine learning algorithm will help us understand the nature of Kepler mission data and its use for various machine learning techniques. We plan to increase our data accuracy in the future with further developments. Machine learning is applied in several areas of space astronomy, such as monitoring astronaut health in orbit, intelligently conducting ship repairs, discovering new planets in other galaxies, and other incredible achievements. Self-driving rovers on Mars, exploring medical capabilities, a Planetary spectrum generator, the Robonaut (A Robotic Astronaut), Deep learning planetary navigation, and many other implementations of machine learning in space exploration and future applications have been achieved so far. Machine learning and Artificial Intelligence (AI) are redefining innovation in the field of astronomy, assisting in the discovery of some of the universe's greatest mysteries. Brant Robertson says: \enquote{Astronomy is on the cusp of a new data revolution, and we couldn't have summarised it better.}

\bibliography{references}
\bibliographystyle{aasjournal}

\end{document}